\documentstyle[12pt,epsf,epsfig,amsfonts]{article}

\begin{document}

\title{Non-random walks in monkeys and humans}


\author{Denis Boyer$^{1,2}$, Margaret C. Crofoot$^{3,4,5}$
and Peter D. Walsh$^{6}$}
\date{}
\maketitle
{\it $ $\\
$\ ^1$Instituto de F\'\i sica, Universidad Nacional Aut\'onoma de 
M\'exico, 04510 D.F., Mexico\\
$\ ^2$Centro de Ciencias de la Complejidad,
Universidad Nacional Aut\'onoma de M\'exico, 04510 D.F., Mexico\\
$\ ^3$Smithsonian Tropical Research Institute, 
Apartado Postal 0843-03092, Panam\'a, Rep\'ublica de Panam\'a\\
$\ ^4$Max Planck Institute for Ornithology, Department of Migration 
and Immuno-ecology, 78315 Radolfzell, Germany\\
$\ ^5$Department of Ecology and Evolutionary Biology, 
Princeton University, Princeton, NJ 08544, USA\\
$\ ^6$VaccinApe, 5301 Westbard Circle, Bethesda, MD 20816, USA
}

\vspace{0.5cm}
\newpage

\begin{abstract}
Principles of self-organization play an increasingly central role in models of 
human activity. Notably, individual human displacements exhibit
strongly recurrent patterns that are characterized by scaling laws and
can be mechanistically modelled as self-attracting walks. 
Recurrence is not, however, unique to human displacements. Here we report that 
the mobility patterns of wild capuchin monkeys are not random walks and 
exhibit recurrence 
properties similar to those of cell phone users, suggesting spatial cognition 
mechanisms shared with humans. We also show that the highly uneven 
visitation patterns within monkey home ranges are not entirely self-generated 
but are forced by spatio-temporal habitat 
heterogeneities. If models of human mobility are to become useful tools for 
predictive purposes, 
they will need to consider the interaction between memory and environmental heterogeneities. 
\end{abstract}


\vspace{0.5cm}
{Short title: Non-random walks in monkeys and humans}
\newpage

\section{Introduction}
Individual human displacements have a strong impact on many collective social 
phenomena, such as the spread of epidemics \cite{vespignani,skufca,belik} 
or cultural traits \cite{sneppen}. The availability of quantitative mobility 
data has increased in recent years through the widespread use of global 
positioning systems \cite{levywalkhuman} and the ability to track cell 
phones \cite{gonzalez,song1,song}. These new data show that human displacements 
do not follow memoryless processes, like the well-known Markovian random
walk \cite{colding}, rather, they exhibit ultra-slow diffusion and 
unusual, long lasting recurrence properties due to the tendency of individuals 
to frequently revisit a small number of familiar locations \cite{song1,song}.

Whether unifying principles that govern the movements of humans \cite{song} 
and, more generally, of living organisms \cite{nathan} exist is a hotly 
debated issue. As with the distribution of city sizes \cite{zipflaw}
or the dynamics of individual tasks \cite{barab}, scaling laws prevail in human 
mobility data \cite{levywalkhuman,song,geisel}. In physical systems, scaling 
laws are often the outcome of self-organization principles. 
Self-attracting or reinforced walks \cite{davis,foster} are non-Markovian 
stochastic walks that tend to revisit with higher probability locations visited 
in the past. Recently, a new reinforced walk model with preferential return 
to previously visited sites, in analogy with preferential attachment 
rules used in network science \cite{barabasi}, could reproduce many empirical 
human visitation patterns \cite{song}. However, it is also important to 
quantify the dependence of individual displacements on more complex external 
factors, for instance, transportation \cite{vespignani} and social 
networks \cite{slaw}, or the spatial distribution of facilities or 
resources \cite{newman}.

In this contribution, we show that the scaling laws characterizing recurrence
in human movements are
similar to those exhibited by some foraging animals, which suggests that 
these patterns are not unique to humans and may be generated by a more 
generalized set of cognitive mechanisms. We present evidence that, like humans, other animals 
also live in heterogeneous habitats where some areas are more valuable to 
them than others.
Recurrent patterns of movement, in particular home range behavior, actually
characterize the ranging of a very large number of animal 
species \cite{moorcroftlewis,borger}. Here, we use radio telemetry data to 
show that the mobility and visitation patterns of capuchin monkeys 
({\it Cebus capucinus}) foraging for fruit in a tropical forest are not 
only comparable to human cell phone users qualitatively, but exhibit 
scaling laws that are strikingly similar in both the short and long time 
regimes. In addition, monkeys movements are not 
entirely self-organized but also strongly driven by spatio-temporal 
variations in resource distribution. 

\section{Results}

The data for our analyses come from a study on Barro Colorado Island, Panama, in which 
the movements of four radio-collared capuchin monkeys belonging to different social 
groups were tracked using an automated telemetry system over the course of 
a six month period, from November 2004 to April 2005 
(see \cite{megpnas} for details). The locations for each individual were 
estimated every $\Delta t_0= 10$ minutes and discretized into square 
cells of size 50$\times$50 m, a scale that corresponds roughly to the 
measurement error and was taken as the spatial resolution. Unless indicated, 
we considered all available positions (night and day) in the analyses. 

\subsection{Basic movement properties}

\begin{figure}
\begin{center}
\includegraphics[width=8.cm]{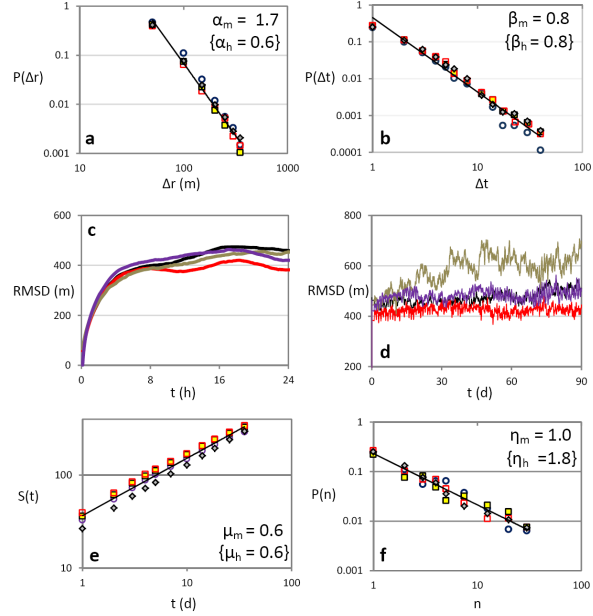}
\caption{Comparison of monkey and human movement properties. 
Data for each of four radio-collared capuchin monkeys plotted separately. 
Monkey power-laws exponents (subscript \lq\lq m") of the frequency
distributions are estimated from pooled data using Bayesian Markov 
Chain Monte Carlo Estimation. These values are practically indistinguishable 
from the MLEs, which are not shown. 
Human power-law exponents from \cite{song} (subscript "h") are given in
each plot for comparison. {\bf a)} Displacements between consecutive 
telemetry fixes ($\Delta r$ in meters) showed a much steeper decay, 
$P(\Delta r) \sim (\Delta r)^{-1-\alpha}$ with larger $\alpha$, 
than in humans but {\bf b)} waiting times (1 unit=10 min)
showed a power-law exponent close to humans. {\bf c)} RMS displacement 
$\langle({\bf r}(t)-{\bf r}(0))^2\rangle^{1/2}$ (in meters) rose quickly during the 
first day but {\bf d)} very slowly thereafter, with values always much smaller 
than the size of Barro Colorado Island (diameter$\sim 5$km). 
{\bf e)} The number of distinct 50x50 m habitat cells visited, $S(t)$, 
as a function of time. {\bf f)} Like for humans, the probability that a cell 
received $n$ visits within the 6-month time interval, $P(n)$, decays as a 
power-law, but with a smaller exponent. 
Estimation cutoffs for $\Delta r$, $\Delta t$ 
and $n$ were set respectively at 350 m, 12 h, and 35 visits.
In all cases the bounded power-law model was a very good fit to the data 
($R^2\ge0.98$) and the frequency distribution with MLE had an Akaike weight 
$>0.9999$ when compared to the most likely exponential distribution. 
The sizes of the $95\%$ confidence intervals of the exponent values were always 
lower than $0.20$.
}  \label{fig1} 
\end{center}
\end{figure}

Short time-scale properties of monkey movement paths 
followed similar scaling forms as those of human cell phone users
Ref. \cite{song} (see Figure \ref{fig1}a,b). For example, the
distributions of the displacement $\Delta r$ between consecutive telemetry 
fixes are good fits to a power-law, $(\Delta r)^{-1-\alpha}$ with
$\alpha\simeq 1.7$ in the range 50 m $<\Delta r<$350 m, 
for all four individuals (sample sizes$=$1619, 1998, 3072 and 2097). 
This exponent, as well as $\beta$ and $\eta$ below, were obtained by using a
Bayesian Markov chain Monte Carlo simulation with uniform priors and
the Metropolis-Hastings algorithm. For each measure we evaluated the 
likelihood of a given value of the exponent of the bounded power-law given 
the data. The exponent value is the average obtained from the posterior 
support distribution. We also calculated the maximum likelihood estimate 
(MLE), obtaining practically the same value.
 
Similarly, the waiting time $\Delta t$, defined as the number of consecutive 
fixes found in the same 50$\times$50 m cell multiplied by the temporal
resolution $\Delta t_0$, follows a power-law distribution, 
$P(\Delta t)\sim (\Delta t)^{-1-\beta}$ with $\beta=0.8$, the 
same value as reported for humans \cite{song,geisel} and other primates 
\cite{gabriel}. The exponents $\alpha<2$ and $\beta\leq 1$ play central roles 
in continuous time random walk (CTRW) models of anomalous 
transport \cite{geisel,metzler} and of biological L\'evy flights 
\cite{gabriel,klafter,viswanathan}. 

Like humans, monkeys trajectories are however incompatible with a random
L\'evy walk description at large time-scales, as they exhibit strongly 
recurrent patterns of movement. For example, the root mean squared displacement 
(RMSD) virtually asymptotes within one day, rather than following a classical 
power-law growth with time (Fig. 1c,d). This arrested or ultra-slow diffusion 
is a manifestation of home range behavior \cite{moorcroftlewis,borger}.
Likewise, the average number $S(t)$ of distinct cells visited during a time 
interval $t$ (Fig. 1e) increases through time as $t^{\mu}$ with $\mu$ smaller 
than the waiting time exponent $\beta$, which is what would have been expected 
for a CTRW \cite{song,metzler}. One finds $\mu=0.6$, remarkably close to what 
measured by Song {\it et al.} for humans \cite{song}. Another intriguing aspect 
of recurrence is the highly uneven visitation pattern among cells, see 
Fig. 1f. Within the six-month data collection period, the probability $P(n)$ 
of finding a cell that has received $n$ visits by the same individual 
({\it i.e.}, has been entered $n$ times) cannot be described
by a Poisson distribution or a bell-curve centered around a characteristic
value. Instead, it can be fitted
by a power-law, $P(n)\sim n^{-\eta}$ with $\eta=1.0$  ($\eta=1.8$ in 
humans \cite{song}). Whereas most sites were visited once or twice, 
it is not rare to find popular activity \lq\lq hotspots" or \lq\lq hubs", 
with 30 or more visits. In contrast, a random walker bounded 
in a closed domain would visit all sites more or less equally frequently. 
For all the properties described above, the four individual monkeys showed 
very similar exponent estimates. The number of distinct
cells visited by the animals during the six-month period was 568, 488, 606 
and 531, respectively.

\subsection{Recurrence and environmental forcings}

Like humans, many animals have sophisticated memory skills \cite{byrne} and 
home ranges can emerge   
from frequent returns to previously visited locations \cite{vanmoorter}.
In principle, home ranges could even be completely self-organized, arising 
as a result of the walker's history. Random walks biased towards the center of mass of 
all previously visited sites, or such that, at each step, there is a finite
probability of returning to a randomly selected site visited in the past,
exhibit arrested or very slow diffusion \cite{borger,gautestad}. 
In the context of human mobility, the results of Figure \ref{fig1} have been 
reproduced assuming that the probability of choosing a given site is 
proportional to the number of previous visits to that site (\lq\lq preferential 
attachment"), and assuming that the probability of taking a random 
step (to a unvisited site) decays algebraically with the number of sites 
already visited, $S(t)$ \cite{song}. An advantage of this latter approach 
is that it builds  up uneven visits and power-laws for the 
distribution $P(n)$.

\begin{figure}
\begin{center}
\includegraphics[width=9.cm,angle=-90]{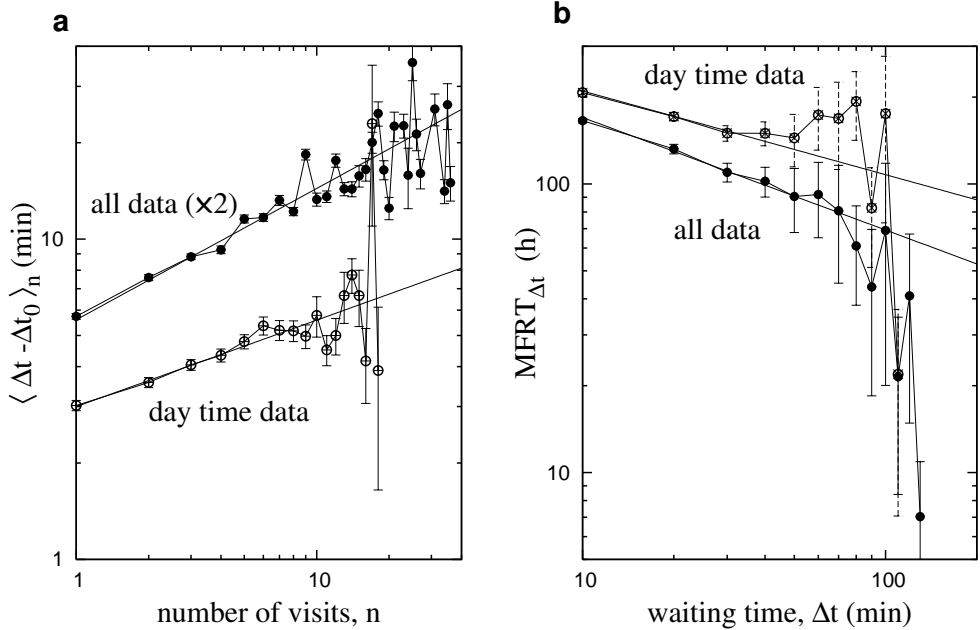}
\caption{{\bf a)} Average visit duration at a cell that has otherwise 
received $n$ visits during a time window corresponding to $S=200$ distinct 
visited sites (about 2 weeks period). Each point represents an average over 
sites with $n$ visits in one time window, over 12 different time windows and 
the 4 different monkeys. The curve obtained from all fixes (\lq\lq all data") 
has been upscaled by a factor 2 for clarity.
{\bf b)} Mean first return time to a cell as a function of the time spent 
on that cell during the last visit. Averages obtained from less than 5 
observations were discarded.
The solid lines are least-square fits 
and have slopes: $0.41$ and $0.27$ in {\bf a)} (all data and day time data,
respectively); 
$-0.39$ and $-0.29$ in {\bf b)} (all data and day time data, resp.). 
In all cases, the probability that the variables are 
uncorrelated ($\delta=\gamma=0$) is $p<10^{-5}$.}  
\label{figfeeding}
\end{center} 
\end{figure}

What these models and those belonging to a much broader class based on preferential 
attachment mechanisms ignore is environmental heterogeneity.
A mobile agent may be inclined to visit a place not because it is familiar
but because of some intrinsic quality of the location, ({\it e.g.}, food content). 
Heterogeneity is implied in the relationship between the quality of 
habitat cells and the probability of cell revisit. Capuchin monkeys home 
ranges typically contain millions of tree stems which belong to more than 
200 species \cite{crofoot}, vary in size by 2 orders of magnitude \cite{enquist} 
and only episodically produce fruit \cite{caillaud}. Consequently, the cells in 
our analyses varied widely in the content of monkeys' primary food, ripe fruit.  
During the day (from 04:00 to 20:00 hrs), animals are awake and 
active \cite{megpnas}, and their primary activities are feeding and foraging
with relatively rare periods of prolonged resting \cite{crofoot}.
During the day, a waiting time of $\Delta t=\Delta t_0=10$ min (one fix,
the minimum observable value) is assumed to correspond to a 
\lq\lq transit" through a cell. Given a series of consecutive fixes
in a same cell, the time elapsed between the
first and the last fixes, $\Delta t-\Delta t_0$, can be 
taken as a proxy of the ripe fruit content within the corresponding 
cell. An individual will tend to spend more time feeding in a tree that 
has more fruit and we consider $\Delta t=\Delta t_0$ as the threshold 
above which one can tell that a feeding event occurred.  
Figure \ref{figfeeding} shows a positive correlation between 
the number $n$ of day-time visits received by a cell and the average duration 
of a visit at that cell, $\langle \Delta t -\Delta t_0 \rangle_n$ 
(the average being taken over all visits among cells with $n$ visits). 
The latter quantity can be approximated by a scaling law:
\begin{equation}\label{feeding}
\langle \Delta t-\Delta t_0\rangle_n\sim n^{\gamma},
\end{equation}
with $\gamma\approx 0.3$. Although other functional relations may fit 
the data, the crucial point is that large $n$ cells have, 
on average, longer waiting times and, most likely, higher food content. 
A similar, and even clearer, relationship is observed if all night and day 
positions are taken into account ($\gamma\approx 0.4$ in this case). 
Sleeping trees also represent resources, as not all trees are suitable 
for sleep at night. 

\begin{figure}
\begin{center}
\includegraphics[width=7.8cm,angle=-90]{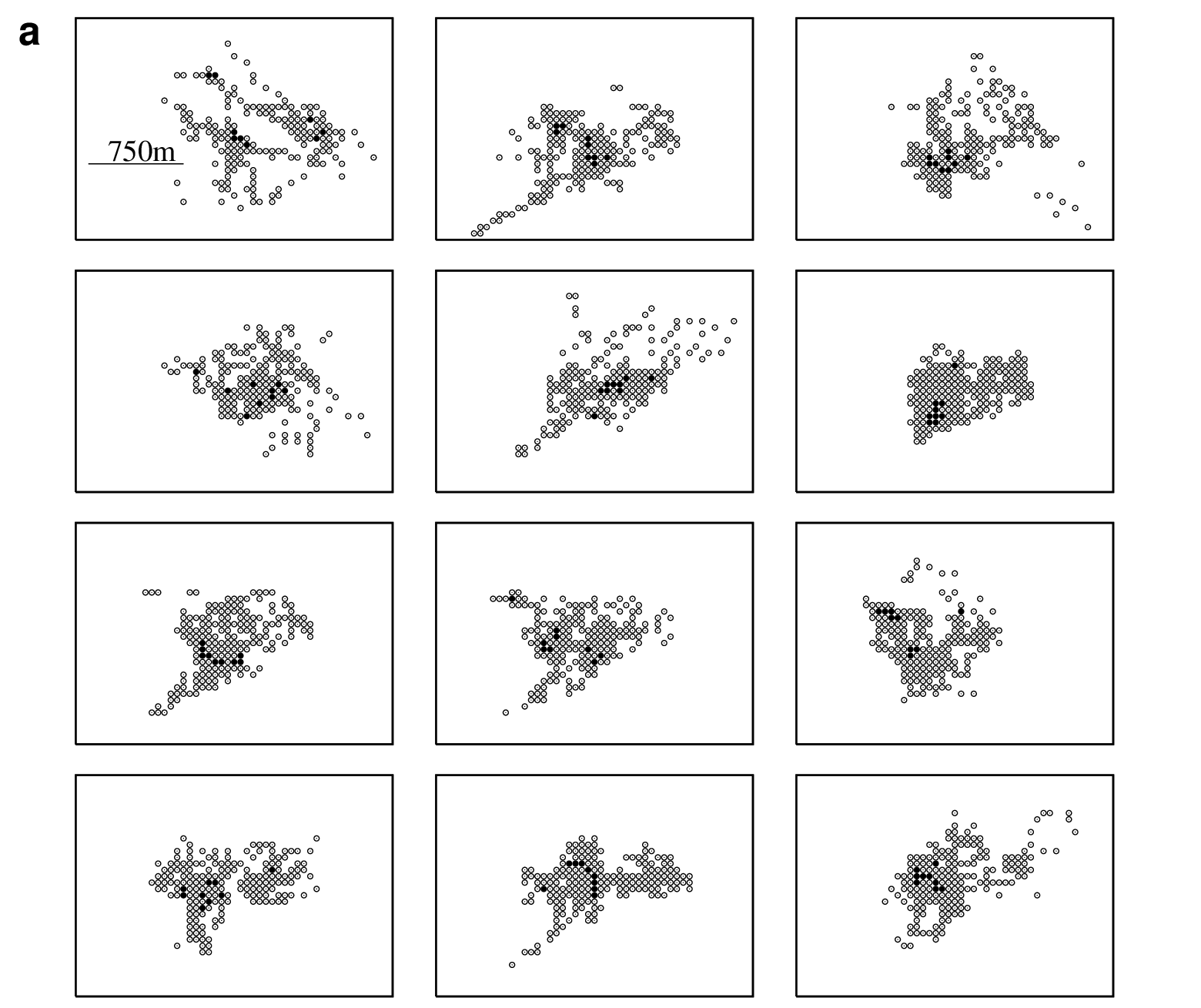}

\includegraphics[width=3.7cm,angle=-90]{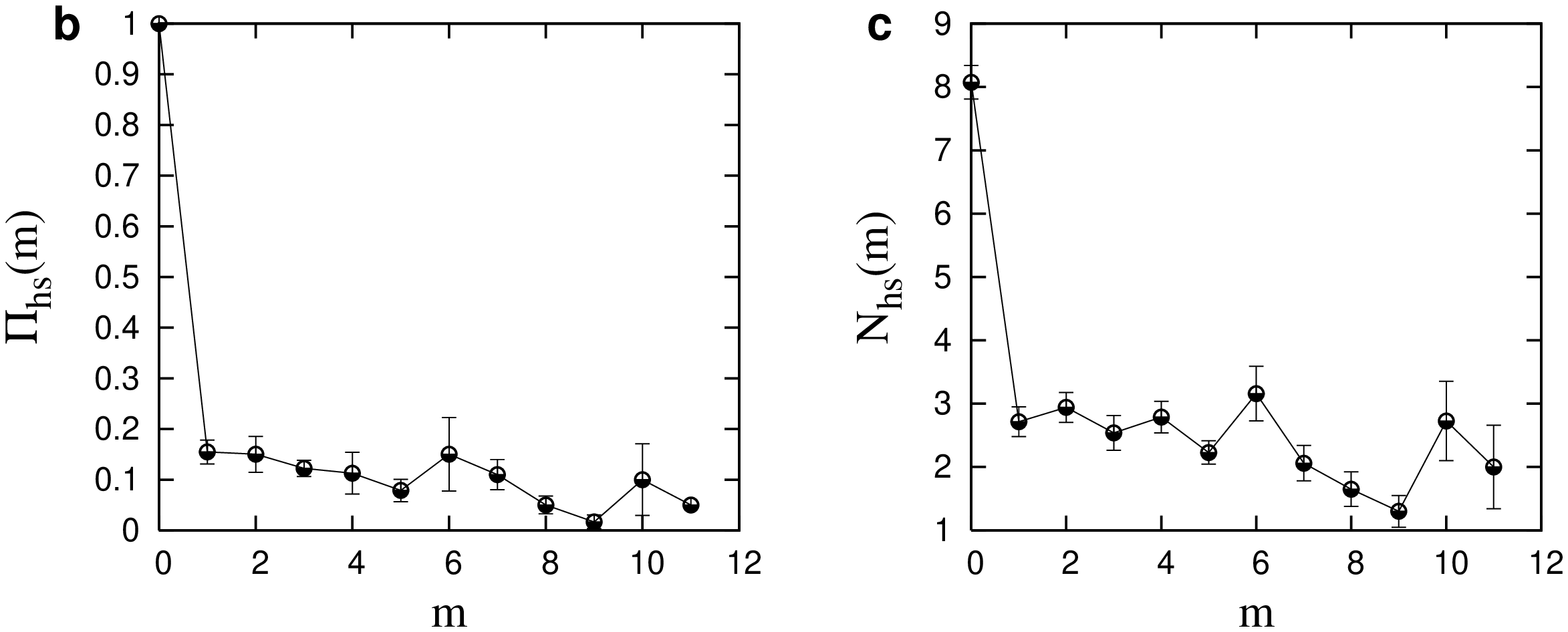}
\caption{{\bf a)} Day-time (4:00-20:00hrs) positions occupied by the same individual monkey 
during successive, non-overlapping periods each containing $S=200$ visited sites. The black dots 
represent the 10 most visited cells during the day, for each time period. Chronological order goes 
from left to right and top to bottom.
Hotspot probability $\Pi_{hs}$ ({\bf b}) and mean number of visits to a hotspot 
$N_{hs}$ ({\bf c}), as a function of the separation $m$ between time windows (see text). 
}  
\label{figmaps} 
\end{center}
\end{figure}

Similarly, the time elapsed between two consecutive day-time visits to a site depends on 
the results of the initial visit. We denote MFRT$_{\Delta t}$ as the mean first 
return time to a cell where the waiting time was $\Delta t$ at the last 
visit. As shown by Figure \ref{figfeeding}b, the MFRT tends to decrease with 
increasing patch quality ($\Delta t$). For $\Delta t\leq 100$ min, it 
approximately follows the scaling law:
\begin{equation}\label{eqmfrt}
{\rm MFRT}_{\Delta t}\sim (\Delta t)^{-\delta},
\end{equation}
with $\delta\approx 0.3$. Note that the unaveraged first return time is a 
variable known to fluctuate widely in simple models like the random 
walk \cite{bouchaud}. Similarly here, error bars in 
Figure \ref{figfeeding}b increase with $\Delta t$, which may be due to 
the fewer available observations at larger $\Delta t$ (Fig. 1b). Averages 
obtained from less than 5 observations were discarded.

These results show evidence that, in addition to being non-random, movement choices 
are driven by resource availability, with larger food patches visited more often.
Eq. (\ref{eqmfrt}) also holds if night and day positions are aggregated 
($\gamma\approx 0.4$).
Hence, a new waiting time is not randomly drawn at each cell visit, as assumed in 
CTRW models. Rather, the situation here is analogous to systems with quenched site 
disorder \cite{bouchaud}. In trap models, though, the frequency of visits to a 
site by a memoryless random walker is independent of the waiting time 
($\gamma=\delta=0$).

Unlike supermarkets, gas stations or other human facilities, resources in natural 
ecosystems tend to be ephemeral. 
Trees produce fruit transiently and often asynchronously, with the fruiting periods of 
individual trees lasting from a few weeks to a couple of months \cite{caillaud}.
Capuchin monkeys hotspots, in particular, are not permanent. Figure \ref{figmaps}a shows 
the positions of the $5\%$ most visited sites by a single individual during the day
in successive non-overlapping time periods (of about 2 weeks) with $S=200$ visited sites each. 
These cells are rarely the same from one time window to the other. The 
fluctuating cloud formed by the visited cells suggests a flexible use of space, 
despite the stability of the overall home range. Defining hotspots as the most
frequently visited $10\%$ of sites, the probability $\Pi_{hs}(m)$ that a cell which 
is a hotspot during a time window $i$ is also a hotspot during time window $i+m$ decays 
rapidly with $m$ (Figure \ref{figmaps}b). The average number of visits 
$N_{hs}(m)$ received by a cell in time window $i+m$, given that this cell was a hotspot 
in time window $i$, also decays with $m$ (Figure \ref{figmaps}c). Very similar curves
are obtained if night and day data are aggregated.

\section{Conclusions}

We have shown that the individual movement patterns of humans are similar 
to those of capuchin monkeys. Most notably these patterns exhibit
ultra-slow diffusion (the presence of a home range) and history dependent recurrence 
properties obeying scaling laws. In monkeys, we have presented evidence 
that these nonrandom walks are caused by the use of memory, 
a behavior which is not unique to humans nor evolutionary novel. 
These results suggest, but do not prove, that similar cognitive mechanisms
may govern the movements of many animals. Testing such hypothesis would 
require data analysis for a variety of species.

Our analyses also indicate that monkey movements are driven by environmental 
heterogeneities. 
Monkey ranging patterns are thus not \lq\lq self-quenched" into a routine 
emerging from initially random movements that would be re-inforced and 
dominated at large time by frequent revisits to a small number of known 
locations, a mechanism proposed for human mobility \cite{song}. Similarly, it is
likely that human movements are not entirely self-organized, as this would 
imply that the locations of activity hotspots would be uncorrelated with 
environmental factors. In the real world, many hotspots would clearly be 
shared because of some intrinsic property of the location: {\it e.g.}, schools, 
transit hubs or office buildings. A difference is that many human resources 
last over long time scales, like homes and workplaces \cite{gonzalez}. Yet, 
others are transient, like restaurants, trendy night-clubs or fashionable 
shops, or seasonal, like swimming pools and ski-resorts. A place can be also 
abandoned for another not because it is depleted but because the appetite 
for that resource has been temporarily sated.

Despite the similarities between human and monkey movements, 
there are important differences. In humans, the home range
size fluctuates widely from one individual to another, ranging
between 1 km and 1000 km approximately \cite{gonzalez,song}. In contrast, 
the nearly identical home range sizes found for the four monkeys 
(Fig.\ref{fig1}c,d) suggest a narrow size distribution, probably due to 
comparable habitats, energetic needs and locomotor capacities. This is 
consistent with the fact that the home range 
of a capuchin group overlaps with a relatively small number of other
home ranges \cite{megpnas}, which probably limits intergroup interactions 
to nearest neighbors or next-nearest neighbors. This
property is not shared by all animal species, though. Territorial prides 
of Serengeti lions occasionally make long-range contacts with other prides 
as well as with nomadic individuals, producing a denser network of contacts 
with the \lq\lq small world" property that characterizes many human social 
networks \cite{lauren}.

Similar to random diffusion models of animal movements with intermittent central 
attraction \cite{blackwell}, several models of human mobility with preferential 
return to home have been introduced recently, showing that recurrence can have dramatic
effects on spreading processes in large populations \cite{vespignani,skufca,belik}. 
These latter models typically assume a known set of locations that can be visited and 
Markovian, individual-dependent transition rates between locations.
Developing parallel modeling frameworks that adequately represent both 
self-organization and environmental forcing will be critical to the success 
of human movement models for purposes such as of controlling the spread of 
infectious diseases. These alternate models should incorporate cognitive 
mechanisms that many vertebrates use, {\it e.g.} spatial representation 
mechanisms (cognitive maps or travel cost discounting) and temporal 
mechanisms (episodic memory) \cite{boyerwalsh}. How scaling laws emerge from 
the interplay between memory and landscape features remains elusive.

\vspace{1cm}

{\bf Acknowledgements}

\vspace{0.3cm}

This work was conducted as a part of the \lq\lq Efficient Wildlife Disease Control: 
from Social Network Self-organization to Optimal Vaccination" Working Group 
supported by the National Center for Ecological Analysis and Synthesis, a centre 
funded by the NSF (grant $\#$DEB-0553768), the University of California, Santa Barbara, 
and the State of California. Funding for the data collection was provided to MCC by 
the Smithsonian Tropical Research Institute and the Frederik Sheldon Travelling 
Fellowship from the Harvard University. Required permits were obtained from the 
Autoridad Nacional del Ambiente, Panama and the Smithsonian Tropical Research 
Institute Institutional Animal Care and Use Committee.  All research complied with 
the laws of the Republic of Panama and the United States.

\label{lastpage}
\end{document}